# Panaches horizontaux non-Boussinesq en milieu homogène

Abdallah Daddi Moussa Ider[1†], Bouzid Benkoussas[1],
Rabah Mehaddi[2], Olivier Vauquelin[2], Fabien Candelier[2]
[1]Laboratoire de génie mécanique et développement, Ecole Nationale Polytechnique.
10 avenue Hassen Badi, 16200 El-Harrach, Alger
[2]Laboratoire IUSTI, UMR CNRS 7343, Aix-Marseille Université
5 rue Enrico Fermi, 13 453 Marseille cedex 13, France.
† **abdallah.daddi-moussa-ider@etu.univ-amu.fr**

**Résumé** : Dans cette étude, nous traitons le problème des panaches ronds éjectés horizontalement dans un milieu homogène au repos. La prédiction du comportement du panache, i.e. l'évolution de ses variables, est d'abord traitée de façon théorique à partir d'un modèle dont le formalisme est valable aussi bien dans le cas Boussinesq que dans le cas général non-Boussinesq. La résolution des équations gouvernant ces panaches est effectuée numériquement à l'aide d'un schéma de Runge-Kutta d'ordre 4. Pour valider le modèle, par rapport à la physique qu'il prétend prédire, des expériences sont réalisées avec des jets ronds d'air et d'hélium pour une large gamme de masses volumiques. La confrontation théorie-expérience vise ici à fixer les limites de la validité du modèle théorique.

**Mots clés** : Panache, non-Boussinesq, tomographie laser, traitement d'image.

## 1. Introduction

La dispersion d'effluents dans des milieux plus denses est un phénomène physique très présent dans la nature. Il peut être observé lors de l'éjection des polluants gazeux par les cheminés d'usines ou lors de l'évacuation des eaux usées issues des processus industriels. Une bonne compréhension et une modélisation simple de ce type d'écoulement est donc nécessaire.

Les paramètres importants sont ici, d'une part, le nombre de Froude qui traduit la compétition entre les forces d'inertie et les forces de flottabilité du rejet, d'autre part, le nombre de Reynolds. Nous considérons dans ce travail que le nombre de Reynolds est suffisamment grand pour s'assurer que l'écoulement est turbulent. Dans ces conditions, la détermination des propriétés du panache ne sera pas affectée par les propriétés moléculaires du fluide. Nous supposons par ailleurs que la vitesse d'entrainement (du fluide ambiant) est proportionnelle à la vitesse locale du panache, par l'intermédiaire d'un coefficient d'entrainement constant, tel que proposé par Morton et al. (1956). Enfin, les profils de vitesses et de masses volumiques sont de type 'top-hat', i.e. qu'ils sont considérés comme étant constants dans une section droite donnée du panache.

Beaucoup d'études ont été réalisées pour étudier les panaches verticaux. Dans le cas des panaches horizontaux ou inclinés, la littérature est plus pauvre. On peut néanmoins citer les travaux de Lane-Serff et al. (1993), Jirka (2004), Kikkert (2006) et Yannopoulos & Bloutsos (2012), basés sur des comparaisons théorie-expérience, mais toujours dans le cadre restrictif de l'approximation de Boussinesq. Michaux & Vauquelin (2008), dans le cas des panaches verticaux, ont introduit un formalisme mathématique permettant de traiter le cas général non-Boussinesq. C'est ce formalisme que nous utiliserons par la suite pour établir un modèle de panache, que nous confronterons ensuite à des expériences de laboratoire.

## 2. Modèle théorique

Comme indiqué sur la figure 1, un fluide léger est éjecté depuis un orifice circulaire de rayon $b_i$, suivant un angle $\theta_i$, et à la vitesse initiale $u_i$ avec une masse volumique initiale $\rho_i$ dans un milieu de masse volumique $\rho_\infty$. Le panache se développe axisymétriquement autour de son axe central repéré par la coordonnée curviligne "s" et par l'angle d'inclinaison θ. Nous considérerons dans cet article que $\theta_i = 0$ mais le modèle analytique proposé sera établi dans un cadre plus général que celui des panaches rejetés horizontalement.

Les équations de conservation pour la masse, la quantité de mouvement (vertical et horizontal) et la flottabilité s'écrivent alors, respectivement, dans un système de coordonnées curvilignes.

$$\frac{d(\beta^2 u)}{ds} = 2\alpha\beta u \qquad (1)$$

$$\frac{d(\beta^2 u^2 \sin\theta)}{ds} = g\eta\beta^2 \qquad (2)$$

$$\frac{d(\beta^2 u^2 \cos\theta)}{ds} = 0 \qquad (3)$$

$$\frac{d(\eta\beta^2 u)}{ds} = 0 \qquad (4)$$

avec $\beta = b(\rho/\rho_\infty)^{\frac{j}{2}}$ le rayon corrigé, $\eta = g\,\Delta\rho/\rho_\infty$ le déficit de masse volumique et $\alpha$ le coefficient d'entrainement. Notons que dans le cas Boussinesq : $j = 0$, et dans le cas général de non-Boussinesq : $j = 1$.

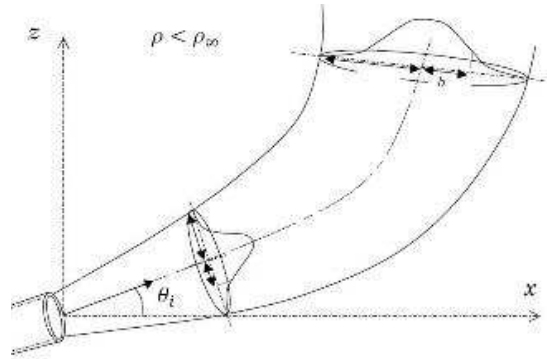

Figure 1 : Schéma descriptif d'un panache éjecté suivant un angle oblique [4].

A partir de ces équations, on peut introduire la fonction panache $\Gamma$ (Michaux et Vauquelin, 2008) que l'on peut interpréter comme étant un nombre de Froude normalisé. Elle est définie par :

$$\Gamma = \frac{5g}{8\alpha}\frac{\eta\beta}{u^2} = \frac{5}{8\alpha Fr^2} \qquad (5)$$

Il est ensuite possible d'exprimer les variables de panache uniquement en fonction de $\Gamma$ et de l'angle $\theta$ :

$$\frac{u}{u_i} = \left(\frac{\Gamma_i}{\Gamma}\right)^{\frac{1}{2}}\left(\frac{\cos\theta}{\cos\theta_i}\right)^{\frac{1}{4}}, \qquad (6\,a)$$

$$\frac{\beta}{\beta_i} = \left(\frac{\Gamma}{\Gamma_i}\right)^{\frac{1}{2}}\left(\frac{\cos\theta_i}{\cos\theta}\right)^{\frac{3}{4}} \qquad (6\,b)$$

$$\frac{\eta}{\eta_i} = \left(\frac{\Gamma_i}{\Gamma}\right)^{\frac{1}{2}}\left(\frac{\cos\theta}{\cos\theta_i}\right)^{\frac{5}{4}} \qquad (6\,c)$$

Finalement, après un peu d'algèbre, on montre que

$$\Gamma(\theta) = \Gamma_i\left(\frac{\cos\theta}{\cos\theta_i}\right)^{\frac{5}{2}} + I(\theta)\cos^{\frac{5}{2}}\theta \qquad \text{avec} \qquad I(\theta) = \frac{5}{2}\int_{\theta_i}^{\theta}\frac{d\theta}{\cos^{\frac{7}{2}}\theta}$$

Le problème se réduit alors à la résolution d'une équation différentielle en $\theta$, malheureusement non-linéaire et n'admettant des solutions analytiques qu'avec une approche basée sur le raccordement de solutions asymptotiques, tel que présentée par Candelier & Vauquelin (2012) pour les panaches verticaux.

Avant de s'attaquer à cette résolution, il nous parait au préalable judicieux de vérifier la validité du modèle défini par les équations de conservation (1, 2, 3, 4). Pour cela, ces équations seront donc résolues numériquement et confrontées à des résultats expérimentaux obtenus en laboratoire sur des panaches non-Boussinesq d'air et d'hélium.

## 3. Etude Expérimentale

### 3.1. Banc d'essai

Les expériences sont réalisées sur un banc d'essai assez sommaire permettant un éclairage par plan laser du panache se développant dans un milieu libre, comme cela est présenté sur la photographie de la figure 2. Un mélange d'air et d'hélium est rejeté par une buse de section circulaire à différents débits. Les débits d'air et d'hélium sont contrôlés par des débitmètres thermiques-massiques Bronkhorst. Pour visualiser l'écoulement, le mélange est ensemencé par des sels de chlorure d'ammonium $NH_4Cl$ obtenus par une réaction chimique entre de l'ammoniaque et des vapeurs d'acide. Les particules sont éclairées par un plan laser RGBLaser de puissance optique de sortie de $2W$, et de longueur d'onde $532\ nm$ (lumière verte). Grâce à la symétrie du problème suivant l'axe perpendiculaire au plan laser, ce plan coïncide parfaitement avec le plan de la trajectoire de la ligne centrale du panache. Les expériences sont enregistrées à l'aide d'une caméra digitale PowerView$^{TM}$Plus, située perpendiculairement au plan laser.

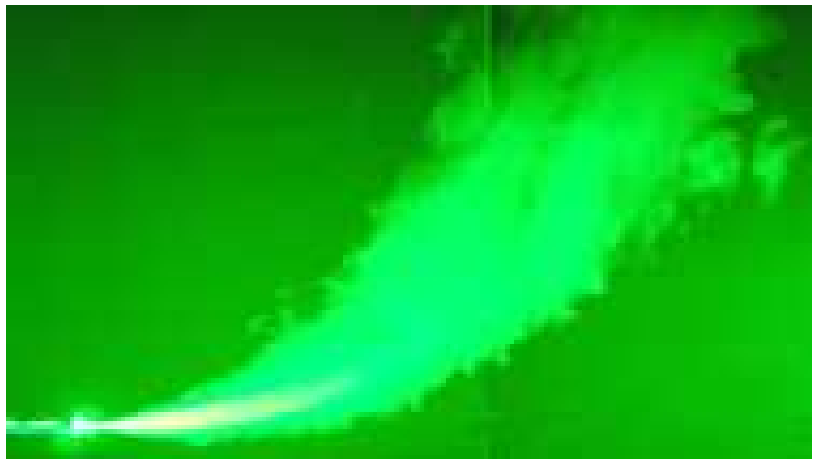

Figure 2 : Photographie d'une expérience.

Ces expériences nécessitent des conditions expérimentales bien contrôlées (pas de perturbations extérieures). Aussi, des parois en plexiglas ont été aménagées autour du banc pour prévenir de ces perturbations, sans altérer néanmoins le processus d'entrainement de l'air ambiant. La fréquence d'acquisition des images, réglable grâce au logiciel CamWare, été fixée à 25 images par seconde. Pour chaque test, nous récupérons donc sur une durée de 30 secondes un peu plus de 600 images. Pour avoir la correspondance entre le pixel de l'image et la longueur réelle, nous utilisons une mire (règle graduée) positionnée dans le plan du panache. Ce dispositif nous permet de fixer la résolution spatiale qui sera ici d'environ un millimètre.

Dans la suite, nous n'exploiterons que les expériences réalisées avec un diamètre de la buse d'éjection de $12.5\ mm$. Pour ce diamètre, plusieurs débits ont été testés.

### 3.2. Post-traitement des images

La deuxième étape consiste à post-traiter les images afin d'en extraire des informations quantitatives sur la trajectoire et l'épanouissement du panache. Ceci est fait à l'aide des logiciels : ImageJ et MATLAB. L'écoulement étant de nature turbulente, la détermination de la trajectoire ne peut se faire qu'après une moyenne des images permettant de s'affranchir des fluctuations. ImageJ est utilisé pour extraire une image moyenne (en intensité de niveaux de gris) à partir de 300 images instantanées. Cette image moyenne est ensuite exportée vers MATLAB pour tracer les contours des lignes d'égales intensités. La figure 3 présente l'image moyennée ainsi que les trajectoires obtenues théoriquement, pour un panache éjecté horizontalement. Pour le même essai, la figure 4 représente les iso-densités basées sur les iso-niveaux de gris.

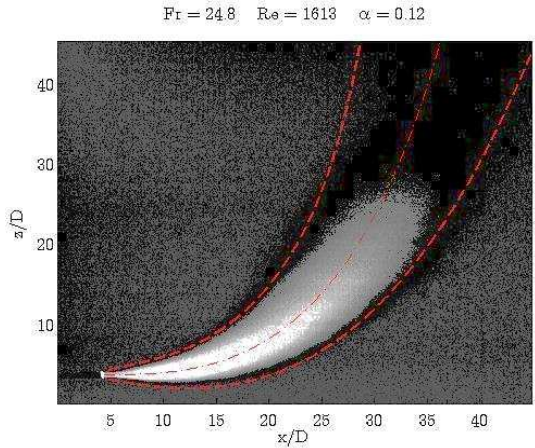 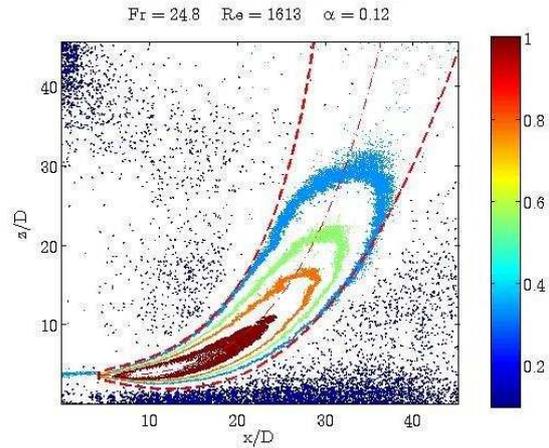

Figure 3 : Image moyennée      Figure 4 : Contours des concentrations

Afin de comparer le modèle théorique aux expériences, nous choisissons comme critère le rayon du panache. En pratique, ce rayon est obtenu, pour les expériences, par une méthode de seuillage des niveaux de gris effectuée avec le logiciel ImageJ. Le seuil du niveau de gris que nous considérons dans cette étude correspond à 35 % du niveau de gris maximal sur toute l'image. Par exemple, la figure 3 présente une image moyenne sur laquelle cette méthode de seuillage a été appliquée. Notons que le maximum de niveau de gris sur l'ensemble des pixels de cette image est de 255. Ceci implique que le seuil de niveaux de gris considéré est de 90. Finalement, tous les pixels ayant un niveau de gris inférieur à 90 seront mis à zéro (zone en noire sur la figure 3). Les zones éclairées quant à elles, correspondent à l'écoulement principal du panache. Quant à la ligne centrale traduisant la trajectoire du panache, elle sera calculée en faisant la moyenne entre l'intrados et l'extrados du panache.

## 4. Discussion et conclusion

Afin de vérifier la validité du modèle donné par les équations (1, 2, 3, 4) nous allons les résoudre numériquement et les confronter aux données expérimentales afin d'évaluer la valeur du coefficient d'entrainement $\alpha$ qui permet d'avoir la meilleur concordance entre la théorie et l'expérience. Le choix de $\alpha$ se fait visuellement de telle sorte que la frontière défini par la méthode de seuillage soit compatible avec la trajectoire déterminée théoriquement. Si ce coefficient conserve toujours une même valeur, on pourra considérer que l'approche théorique est satisfaisante, sinon, en cas de forte dispersion de la valeur de $\alpha$ il conviendra de réfléchir à une à modélisation plus élaborée du processus d'entrainement.

Pour des nombres de Froude $Fr = 24.8, 31.3, 36.8$ et $44.9$ relatifs à chaque expérience, les valeurs du coefficient d'entrainement $\alpha$ sont de $0.12, 0.13, 0.13$ et $0.13$. Les comparaisons entre la théorie et l'expérience sont montrées sur la figure 5. Nous remarquons que le modèle reproduit de façon satisfaisante la trajectoire ainsi que le rayon du panache.

A ce stade, il est clair que le modèle a bien été validé (pour un panache horizontal), malgré le fait que la méthode expérimentale mise en place est extrêmement simple et nous a permis d'obtenir des résultats très rapidement. Néanmoins, nous pensons qu'il est judicieux de reprendre ses expériences avec une méthode optique mieux contrôlée (éclairage par faisceau ou PLIF). En effet, la méthode expérimentale utilisée dans cette étude présente des incertitudes, notamment, liées à la non uniformité de l'intensité lumineuse du plan laser utilisé pour éclairer l'axe médian du panache. A un niveau moindre, le phénomène d'extinction lumineuse au sein des fumées est également une source d'incertitude sur la détermination de la trajectoire. Les résultats encourageants de cette étude devront aussi être élargies aux cas d'inclinaisons différentes ($\theta_i \neq 0$).

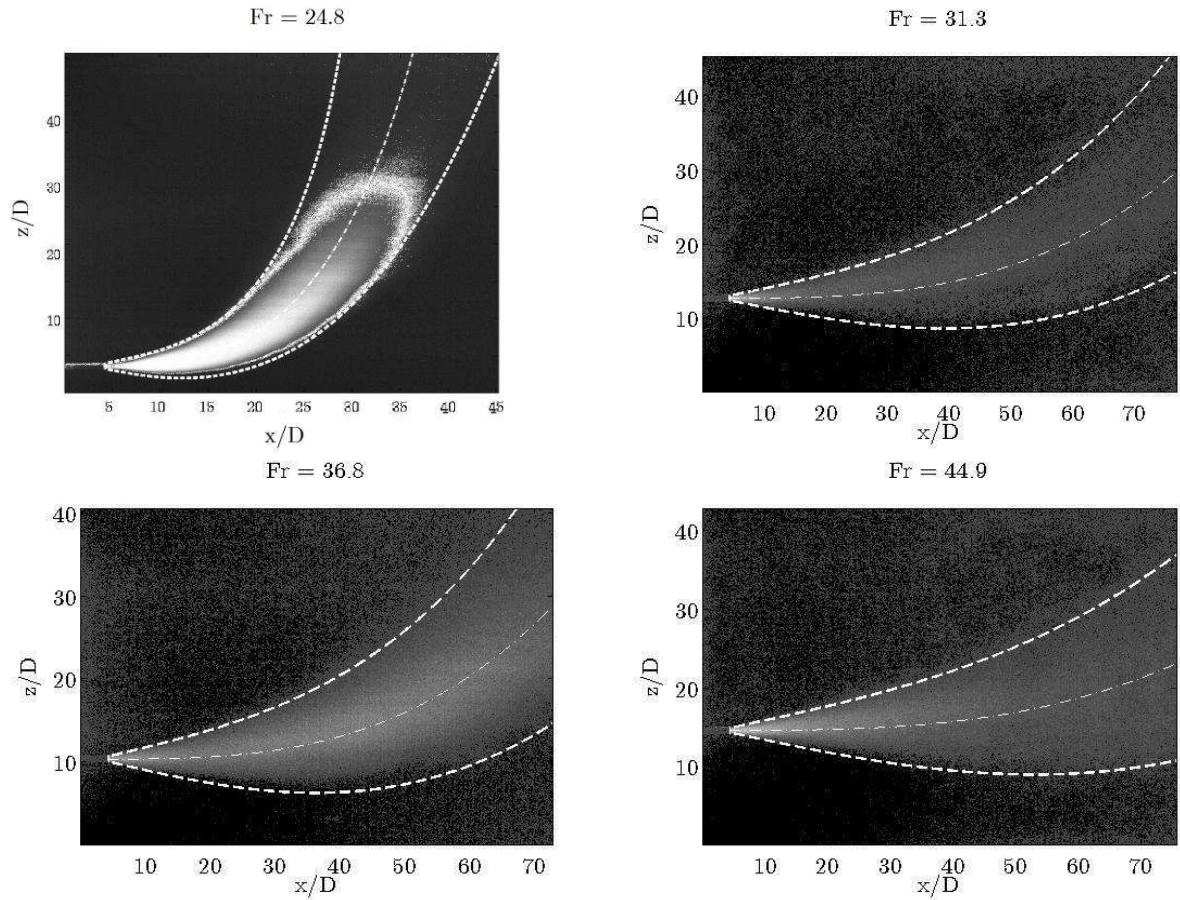

Figure 5 : Seuillage des images moyennées

**Nomenclature**

| Symbole | Nom, unité | | Symboles grecs | |
|---|---|---|---|---|
| u | vitesse, m/s | | $\alpha$ | coefficient d'entrainement |
| s | coordonnée curviligne, m | | $\beta$ | rayon modifié, m |
| Fr | nombre de Froude | | $\theta$ | angle d'inclinaison |
| | | | $\eta$ | déficit de densité |
| Exposant, Indices | | | $\rho$ | densité du panache, kg.m$^{-3}$ |
| | | | $\rho_\infty$ | densité du milieu ambiant, kg.m$^{-3}$ |
| i | interne | | $\Delta\rho$ | écart des densités, kg.m$^{-3}$ |
| | | | $\Gamma$ | fonction panache |